\documentclass[10pt,journal,compsoc]{IEEEtran}

\ifCLASSOPTIONcompsoc
\else
\fi

\ifCLASSINFOpdf

\else

\fi

\usepackage{graphics} 
\usepackage{epsfig} 
\usepackage{amsmath} 
\usepackage{amssymb}  
 \usepackage{csquotes}
\usepackage{bm}
\usepackage{amsmath}
\usepackage{url}
\usepackage{gensymb}
\usepackage[utf8]{inputenc}
\usepackage[T1]{fontenc}
\usepackage{textcomp}
\usepackage{gensymb}
\usepackage{ragged2e}
\usepackage{algorithm}
\usepackage{algpseudocode}
\usepackage{color}
\usepackage{multirow}
\usepackage[noadjust]{cite}
\begin{document}

\title{A Joint Indoor WLAN Localization and Outlier Detection Scheme Using LASSO and Elastic-Net Optimization Techniques}

\author{Ali~Khalajmehrabadi\IEEEauthorrefmark{1},~\IEEEmembership{Student~Member,~IEEE,}
        Nikolaos~Gatsis\IEEEauthorrefmark{1},~\IEEEmembership{Member,~IEEE,}\\
        Daniel~Pack\IEEEauthorrefmark{2},~\IEEEmembership{Senior Member,~IEEE,}
        and~David~Akopian\IEEEauthorrefmark{1},~\IEEEmembership{Senior~Member,~IEEE}
\thanks{Manuscript submitted on January, 3, 2016, revised on April, 10, 2016, July, 20, 2016, and September, 23, 2016. Accepted on September, 24, 2016. }
\thanks{\IEEEauthorrefmark{1}The authors are with the Department of Electrical and Computer Engineering, the University of Texas at San Antonio, One UTSA Circle, San Antonio, Texas 78249-0669, USA, E-mails: ali.khalajmehrabadi@utsa.edu, nikolaos.gatsis@utsa.edu, david.akopian@utsa.edu.}

\thanks{\IEEEauthorrefmark{2}The author is with the department of Electrical Engineering, The University of Tennessee at Chattanooga,  615 McCallie Ave, E-mail: daniel-pack@utc.edu.}
}
\markboth{IEEE Transactions on Mobile Computing  (ACCEPTED)}
{}

\IEEEtitleabstractindextext{%
\justify \begin{abstract}In this paper, we introduce two indoor Wireless Local Area Network (WLAN) positioning methods using augmented sparse recovery algorithms. These schemes render a sparse user's position vector, and in parallel, minimize the distance between the online measurement and radio map. The overall localization scheme for both methods consists of three steps: 1) coarse localization, obtained from comparing the online measurements with clustered radio map. A novel graph-based method is proposed to cluster the offline fingerprints. In the online phase, a Region Of Interest (ROI) is selected within which we search for the user's location; 2) Access Point (AP) selection; and 3) fine localization through the novel sparse recovery algorithms. Since the online measurements are subject to inordinate measurement readings, called outliers, the sparse recovery methods are modified in order to jointly estimate the outliers and user's position vector. The outlier detection procedure identifies the APs whose readings are either not available or erroneous. The proposed localization methods have been tested with Received Signal Strength (RSS) measurements in a typical office environment and the results show that they can localize the user with significantly high accuracy and resolution which is superior to the results from competing  WLAN fingerprinting localization methods.
\end{abstract}

\begin{IEEEkeywords}
 Indoor positioning, WLAN fingerprinting, sparse recovery, outlier detection
\end{IEEEkeywords}}

\maketitle

\IEEEdisplaynontitleabstractindextext

\IEEEpeerreviewmaketitle

\IEEEraisesectionheading{\section{Introduction}\label{sec:introduction}}

\IEEEPARstart{R}{ecent} communication systems increasingly rely on Location-Based Services (LBSs) \cite{1} to provide mobile devices with more efficient content delivery such as network management and security \cite{2}, \cite{3}, emergency personnel navigation, healthcare monitoring \cite{4}, personalized information delivery \cite{5}, context awareness \cite{6}, etc. While LBS has been established in outdoor environments through the Global Positioning System (GPS) \cite{7} and become widespread in the last two decades, there is no established infrastructure in indoor areas yet. Although some approaches utilize the GPS indoors, the complicated indoor environment structures prohibited these approaches to succeed. Hence, more sophisticated schemes are necessary to tackle the indoor positioning task. In indoor settings, signals cannot penetrate the metallic obstacles from where multiple copies of a single signal are spread. Also, the propagation environment rapidly changes in vibrant commuting settings. Considering these, robust indoor positioning still remains as an open problem that needs to be investigated.
\par Indoor localization techniques have employed proximity sensors such as RF-tag  \cite{58}, existing RF network infrastructures such as Bluetooth \cite{10,81}, ultrasonic and sound techniques \cite{70,71,72,73,74,80}, visible light \cite{75,76}, motion sensors \cite{77,82}, magnetic field exploitation \cite{78,79} or even infrared transceivers \cite{9}. These methods need a new infrastructure for deploying and maintaining these sensors. 
\par For the last decade, a set of approaches  have utilized Wireless Local Area Network (WLAN) for indoor positioning. Although this infrastructure has been primarily designed and established for an optimal network coverage, it can also deliver quite unique location-specific profile, based on which the user's position can be estimated. From the network side, this approach minimizes the system infrastructure cost; and from the user's perspective, there is no need to hold any specific device other than the one that connects to the wireless network. WLAN positioning techniques are typically categorized into three groups based on their processing type of WLAN signals: (1) Angle of Arrival (AOA) \cite{11,12,13,65,68}; (2) Time of Arrival (TOA) or Time Difference of Arrival (TDOA) \cite{14,15,16,17,66,67}; and (3) fingerprinting methods \cite{18,19,20,21,r39,r27,r33,54,55}.
\subsection{Overview of WLAN Fingerprinting Localization}
\par Amongst these, fingerprinting has gathered much attention lately.   
In more detail, fingerprinting comprises two phases. In the \textit{offline phase}, the area is segmented into reference grids, also called landmarks or Reference Points (RPs). Then, at each RP, typically Received Signal Strength (RSS) profile is recored from the visible WLAN Access Points (APs), whose locations are not necessarily known. These RSS profiles are quite unique and are also called signal fingerprint (signature) of that specific location (one should also note that other signal characteristics can be used along with RSS measurements for the fingerprints). The whole set of RSS fingerprints for all RPs is called the \textit{radio map} of that area. For positioning, in the \textit{online phase}, the user with unknown location receives the RSS measurements from a set of APs. The online measurement is compared to the database of radio map through which the user location is estimated. This comparison method is called a rule or positioning algorithm.  
\par Recent localization schemes perform three tasks in the online phase for increasing the accuracy and decreasing the computational complexity. These tasks include: 1) AP selection, in which a subset of APs that provide more stable data and better differentiate between RPs is selected; 2) coarse localization, whereby the positioning scheme selects a subset of RPs (sub-area) in the vicinity of the user; and 3) fine localization which introduce an efficient metric that is used  to find the closest RPs to the user. We briefly discuss these tasks in the prior art in the following subsection.

\subsection{Related Works in Fingerprinting Localization}
\par In general, two main approaches are employed in WLAN-fingerprinting based approaches \cite{85,86}: deterministic methods such as RADAR \cite{8} and probabilistic \cite{59} such as Horus \cite{83}. Deterministic methods rely on  metrics between online and representative offline fingerprints, while probabilistic techniques exploit statistical likelihoods of online measurements at different locations, and are based  on  the  Maximum a Posteriori (MAP) or Maximum Likelihood (ML) criteria. 
\par Pioneering deterministic approaches benefit from the k-Nearest Neighbor (kNN) criterion which utilizes the Euclidean distance between the online measurements and radio map fingerprints. The estimated location is in the convex hull of the k RPs with the least distances \cite{8,84,85} and weights can also be assigned to each RP based on the  similarity between the online measurement and each fingerprint Weighted KNN (WKNN) \cite{101}. Tilejunction \cite{96,100}, Contour-based trilateration \cite{99}, and Sectjunction \cite{97} are also recent methods that formulate the localization problem into a convex program (with linear \cite{96,100,99} and  quadratic \cite{97}  environmental constraints such as the presence of walls, allowed area, etc.)
\par The work in \cite{18} is one of the earliest contributions in probabilistic WLAN positioning algorithms, in which a subset of the strongest APs is chosen to provide a constant coverage over the area with a higher probability. Kushki et al. \cite{19} proposed a kernelized-based positioning scheme. In this work, RPs are clustered using AP indicator vector which compares the difference between the online measurement and  radio map fingerprints. Several approaches such as strongest AP, Bhattacharyya distance, and Information Potential (IP) have been employed to select the suitable subset of APs. The estimated position is a weighted average of the selected points, where these weights are obtained using an inner product between the online and radio map fingerprints using a Kernel equivalent function. More sophisticated probabilistic techniques have been recently investigated  including Kullback-Leibler (KL) divergence \cite{87}, Principal Component Analysis (PCA) \cite{88}, Conditional Random Fields (CRF) \cite{98} and Bayesian Networks \cite{89}. 
\par Probabilistic methods guarantee higher localization accuracy as they exploit more accurate statistical representation of RSS measurements. However, their computational complexity increases significantly as well.
\par There exists trend of indoor localization enhancement utilizing assistance data from other available sensors which are deployed in commonly used wireless devices, such as ambient lights/colors, ambient sounds \cite{90}, acoustic ranging  Centaur \cite{61}, RSSI from nearby base stations \cite{91}, RFID tags \cite{92}, mostly for a proximity determination of the user so that the localization algorithm is solved in a smaller area. The EZ method  \cite{60} reduces the surveying burden and achieves better accuracy compared to other model-based approaches, however, experiences inferior location accuracy compared to fingerprinting techniques such as  RADAR \cite{8} and Horus \cite{83}, and needs GPS fixes to remove the location ambiguities.  The premise is proximity determination of the user so that the localization algorithm is solved in a smaller area \cite{86}. In addition, motion-assisted localization exploits inertial  accelerometers, gyroscopes, and magnetometers \cite{95} are also very common. These sensors are employed for smoother trajectory estimation, and to determine the walking direction \cite{93}, walking detection status, and step counting of the user \cite{94} such as LiFS \cite{62} etc.. The information from these senors can either be exploited to estimate the user's location independently (with finally fusing the location estimate with that of RSS fingerprints) or directly assisting in the location estimation with RSS fingerprints.
\par While these approaches are valuable contributions to the state-of-the-art, this paper puts forth essential enhancement using only WLAN fingerprints, which can be further improved by integrating supplemental assisted data similar to  \cite{60,61,62}.
\par Recently new fingerprinting approaches are proposed by leveraging sparse signal processing techniques \cite{r39} and \cite{r27}. These methods can be categorized as a new era in deterministic approaches. The estimated position indicator vector can be considered as a vector where only one or a small subset of indices are nonzero. The WLAN positioning problem can therefore be cast as finding a sparse position vector based on the online measurements and radio map fingerprints. This vector can be estimated via solving an $  \ell_{1} $-minimization problem, which amounts to the Compressive Sensing (CS) approach. The details will be discussed later as a basis for indoor CS WLAN localization. The RSS Signal Strength Difference (SSD) has also been used for CS-based positioning \cite{r33}.

\subsection{Proposed Localization System Framework}
\par In this paper, RSS fingerprints are recorded in four orientations in the offline phase. Then, a clustering scheme is applied on the radio map so that the RPs with the most similarity are categorized in a cluster. One RP becomes the representative of its cluster, called \textit{Cluster Head} (CH).  In online phase, we compute the similarity between the online measurement and that of all CHs and the cluster corresponding with the least distance is selected as the coarse location of the user. Hence, the localization is confined to a subset of RPs, called the Region Of Interest (ROI).  This leads to a subset of radio map called modified radio map.
\par Unlike the conventional localization schemes which select the APs based on the fingerprints, we apply the AP selection during the online localization phase so that the APs that better suit the user's location are selected. We utilize the known Fisher criterion as an optimum AP selection method since it takes into account both the time and spatial variations of RSS fingerprints.
\par For fine localization, a novel framework containing a class of formulations for sparse user location recovery is proposed. We reformulate the fine localization problem into two optimization algorithms. One, known as Least Absolute Shrinkage and Selection Operator (LASSO) \cite{63}, seeks for a sparse position vector which satisfies two criteria: (1) The position vector is sparse so that only the most similar RPs give weight to the estimated pose; (2) The difference between the radio map and online measurements is minimized as the sparsity is not sufficient and the selected RPs' fingerprints should be the most similar to the online measurement. Also, since the resultant modified radio map may contain fingerprints in different orientations from a reference point, it is prone to have similar, partially correlated entries. The second localization formulation, known as Elastic-net Regularized Generalized Linear Models (GLMNET) \cite{64}, includes the previous two criteria of the LASSO. However, GLMNET averages the correlated predictors and enter the averaged predictor into the model.  These methods have been previously used in cancer data analysis, logistic regression and gene selection and showed promising results \cite{63,64}.
\par In addition, the online measurements are prone to contain inordinate errors called \textit{outliers}. These outliers are frequent as some APs may not be available during the online localization, or there could be interference potentially introduced by adversary attacks. The online RSS readings from the outlier-ridden APs are highly biased and cause large position estimation errors. This issue has surprisingly received little attention in the  literature, even though it plays an important role in the practical localization systems. Another contribution of this paper is to reformulate the sparse recovery algorithms so that they detect the possible outliers in the online measurements. To this end, we add an outlier detection component to the CS localization and the two proposed schemes, LASSO and GLMNET, so that the outliers and user' location are jointly estimated.
\par This paper discusses a typical WLAN fingerprinting localization and the corresponding Compressive Sensing (CS) problem formulation in Section 2. The proposed localization procedure, consisting of offline clustering, AP selection, and fine localization is elaborated in Section 3. The joint localization and outlier detection schemes are discussed in Section 4. Section 5 illustrates the experimental performance of our methods in a real environment followed by conclusions in Section 6. 
\section{WLAN Fingerprinting Localization: Definitions and Formulation}
In this section, we first introduce the general WLAN Definitions in Sec. \ref{Definitions}, and then Sec. \ref{Measurement Model for Sparse Recovery} gives the measurement model enabling sparse recovery. Afterwards, we discuss the CS formulation and its shortcomings in Sec. \ref{Shortcomings}.

\subsection{Definitions} 
\label{Definitions}
This section provides an ideal indoor WLAN fingerprinting localization problem formulation and a description of the conventional methods. 
In fingerprinting, the area is divided into a set of RPs   $ \mathcal{P}=\left\{\mathbf{p}_{j}=(x_{j},y_{j})| j=1,\ldots,N  \right\} $ where $  \mathcal{P} $ defines the set of RP Cartesian coordinates, which are not necessarily set apart with equal distances. At each RP, the mobile device records the RSS fingerprints at time instants $t_{m},\ m=1,\ldots,M$ with recorded RSS magnitudes $ r_{j}^{i,o}(t_{1}),r_{j}^{i,o}(t_{2}),\ldots, r_{j}^{i,o}(t_{M}) $, where $ i $ indicates the AP index from the set of APs $\mathcal{A}= \left\{AP^{1},AP^{2},\ldots,AP^{L}\right\} $ at orientation $o \in \mathcal{O}= \{ 0\degree ,90\degree ,180\degree ,270\degree \}$. It is typical to take the same number of training samples, $ M $, at each RP. The RSS fingerprints from all APs at $\mathbf{p}_{j}$ and at time $t_{m}$ are organized in a vector $ \mathbf{r}_{j}^{o}(t_{m})=[r_{j}^{1,o}(t_{m}),r_{j}^{2,o}(t_{m}),\ldots,r_{j}^{L,o}(t_{m})]^{T} $. The entire radio map at recording instant $ t_{m} $ can be represented as
\begin{equation}
\begin{split}
 \mathbf{R}^{o}&(t_{m})=\left[  \mathbf{r}_{1}^{o}(t_{m}),\mathbf{r}_{2}^{o}(t_{m}),\ldots,\mathbf{r}_{N}^{o}(t_{m})\right]   =\\
 &\begin{pmatrix}
   r_{1}^{1,o}(t_{m}) & r_{2}^{1,o}(t_{m})  & \cdots & r_{N}^{1,o}(t_{m}) \\
   r_{1}^{2,o}(t_{m}) & r_{2}^{2,o}(t_{m})  & \cdots & r_{N}^{2,o}(t_{m}) \\
   \vdots  & \vdots  & \ddots & \vdots  \\
   r_{1}^{L,o}(t_{m}) & r_{2}^{L,o}(t_{m}) & \cdots & r_{N}^{L,o}(t_{m}) 
  \end{pmatrix}\\
  & \ \ \ \ \ \ \ \ \ \ \ \ \ \ \ \ \ \ \ \forall m=1,\ldots,M.
  \end{split}
\end{equation}
If the time sequence of radio maps, $  \mathbf{R}^{o}(t_{m}) $, is averaged over the recording time, the time averaged radio map is denoted as $ \mathbf{\Psi}^{o}=\left[ \boldsymbol {\psi}_{1}^{o},\mathbf{\boldsymbol \psi}_{2}^{o},\ldots,\mathbf{\boldsymbol \psi}_{N}^{o}\right]   $ where $ \boldsymbol \psi_{j}^{o}=[\psi_{j}^{1,o},\psi_{j}^{2,o},\ldots ,\psi_{j}^{L,o}]^{T}$, and $\psi _{j}^{i,o}=\frac{1}{M}\sum_{m=1}^{M} r_{j}^{i,o}(t_{m})$. In later discussion, we use the notation $\mathbf{\Psi}$ which is the modified radio map obtained from $\mathbf{\Psi}^{o}$ by any algorithm. For instance, $\mathbf{\Psi}$ can be an average of $\mathbf{\Psi}^{o}$ over all orientations.
\par In the online phase, the mobile user receives the online RSS measurements, $ \boldsymbol y=\left[ y^{1},y^{2},\ldots,y^{L}\right] ^{T} $.
The goal of a localization scheme is to find the mobile user's location, $  \hat{\mathbf{p}}= (\hat{x},\hat{y})$, based on a measure that compares the received online measurements and radio map fingerprints.
\subsection{Measurement Model for Sparse Recovery}
\label{Measurement Model for Sparse Recovery}
Most of the conventional fingerprinting methods are computationally intensive, however, CS has opened a new door to perform localization in an acceptable processing time and accuracy \cite{r39,r27}. Ideally, assume that the user is present exactly at one of the RPs. Then, the localization problem can be cast as assigning one of the RPs in the area as the user's location. This localization problem has a sparse nature and hence, the position estimation problem can be formulated as finding an indicator pose vector whose only one element is nonzero (1-sparse vector). So, the localization problem can be recast as an $ \ell_{1} $-minimization problem known as CS. 
\par Introduce a sparse location indicator vector  $\boldsymbol\theta = [ 0,\ldots,0,1,0,\ldots,0 ]^{T} $ where each entry of $ \boldsymbol \theta  $   corresponds to an RP and 1 denotes the index of the RP to which the user is the closest. The measurement model for WLAN localization to enable sparse recovery is
\begin{equation}
\label{eq2}
\mathbf{y}=\mathbf{\Phi \Psi \boldsymbol\theta +\boldsymbol\epsilon}
\end{equation}
where $  \boldsymbol \Phi  $ is the AP selection matrix, i.e. the matrix that selects certain elements of $\mathbf{\Psi}$ corresponding to selected APs, $ \boldsymbol \Psi $ is the modified radio map matrix,$ \ \boldsymbol \epsilon $ is the error vector, and $ \mathbf{ y }$ is the online  captured RSS vector from specific APs as
\begin{equation}
\label{eq3}
\mathbf{y}=\boldsymbol \Phi \boldsymbol y.
\end{equation}

\subsection{CS Formulation and Shortcomings}
\label{Shortcomings}
The CS formulation follows models \eqref{eq2} and \eqref{eq3}  except that it does not take the measurement noise $\boldsymbol\epsilon$ into account. Since the size of $  \mathbf{y} $ is less than that of $ \boldsymbol \theta $, the CS problem is an under-determined problem. However, $ \boldsymbol \theta $ is sparse as the user can only be in one of the RP locations and the problem can be solved via the convex optimization
\begin{equation}
\begin{split}
\label{eq4}
 \hat{\boldsymbol\theta} =\text{argmin} \Arrowvert \boldsymbol\theta \Arrowvert_{1} \\
  s.t.\  \mathbf{y}=\mathbf{\Phi \Psi \boldsymbol\theta}
\end{split}
\end{equation}
where $ \Arrowvert \boldsymbol \theta \Arrowvert _{1} $ is the $\ell_{1}$-norm of $ \boldsymbol\theta $. Using the $ \ell_{1}$-norm, the CS renders a sparse vector. This optimization formulation assumes that the model (\ref{eq2}) does not contain the measurement error $\boldsymbol \epsilon$. Several algorithms have been proposed to solve this problem, e.g. greedy algorithms \cite{22}, \cite{23}, Iteratively Re-weighted linear Least-Squares (IRLS) problems \cite{24}, \cite{25}, basis pursuit \cite{26}, etc. The basic idea of these algorithms is that they tend to converge to a unique solution if specific conditions are satisfied. 

\par  The previous method amounts to the fine localization task, which is paired with an offline clustering scheme in \cite{r27}, and has several shortcomings. The offline clustering methods that need the number of clusters a priori (such as affinity propagation in \cite{r27}) force the area to be divided in a given number of clusters regardless of the similarities between them. In addition, even if the user resides on the border between two clusters, only one of them will be chosen in the coarse localization. Furthermore, the CS recovery scheme needs the sensing matrix $ \boldsymbol\Phi $ and the basis (dictionary) matrix  $ \boldsymbol\Psi $ to obey two criteria in order to obtain a unique optimal solution: 1) the mutual coherence between $ \boldsymbol\Phi $  and   $ \boldsymbol\Psi $ should be sufficiently small \cite{31}; and 2) the product of sensing and basis matrix, i.e.  $ \boldsymbol\Phi\boldsymbol\Psi $, should be nearly orthogonal \cite{32}. Thus, an orthogonalization procedure is applied to induce the incoherence property as required by the CS theory. Nonetheless, the orthogonalization does not make $ \boldsymbol\Phi\boldsymbol\Psi $ completely orthogonal, as it is not square.

\par The shortcomings of the offline clustering in \cite{r27} are overcome in our approach that utilizes the characteristics of the environment to cluster the RPs. The number of clusters depends on the environment and is not defined \textit{a priori}. In addition, the LASSO-based and GLMNET-based recovery methods do not need the orthogonalization step, and do not rely on special properties of the matrix $ \boldsymbol\Phi\boldsymbol\Psi $, which may not be satisfied in practice. In addition to recovering a sparse vector, the proposed localization methods use (\ref{eq2}) as the model, and thus, work better with noisy measurements.

\section{Novel Localization Approach Based on Sparse Recovery}
\par
\begin{figure}[t!]
      \includegraphics[width=\linewidth]{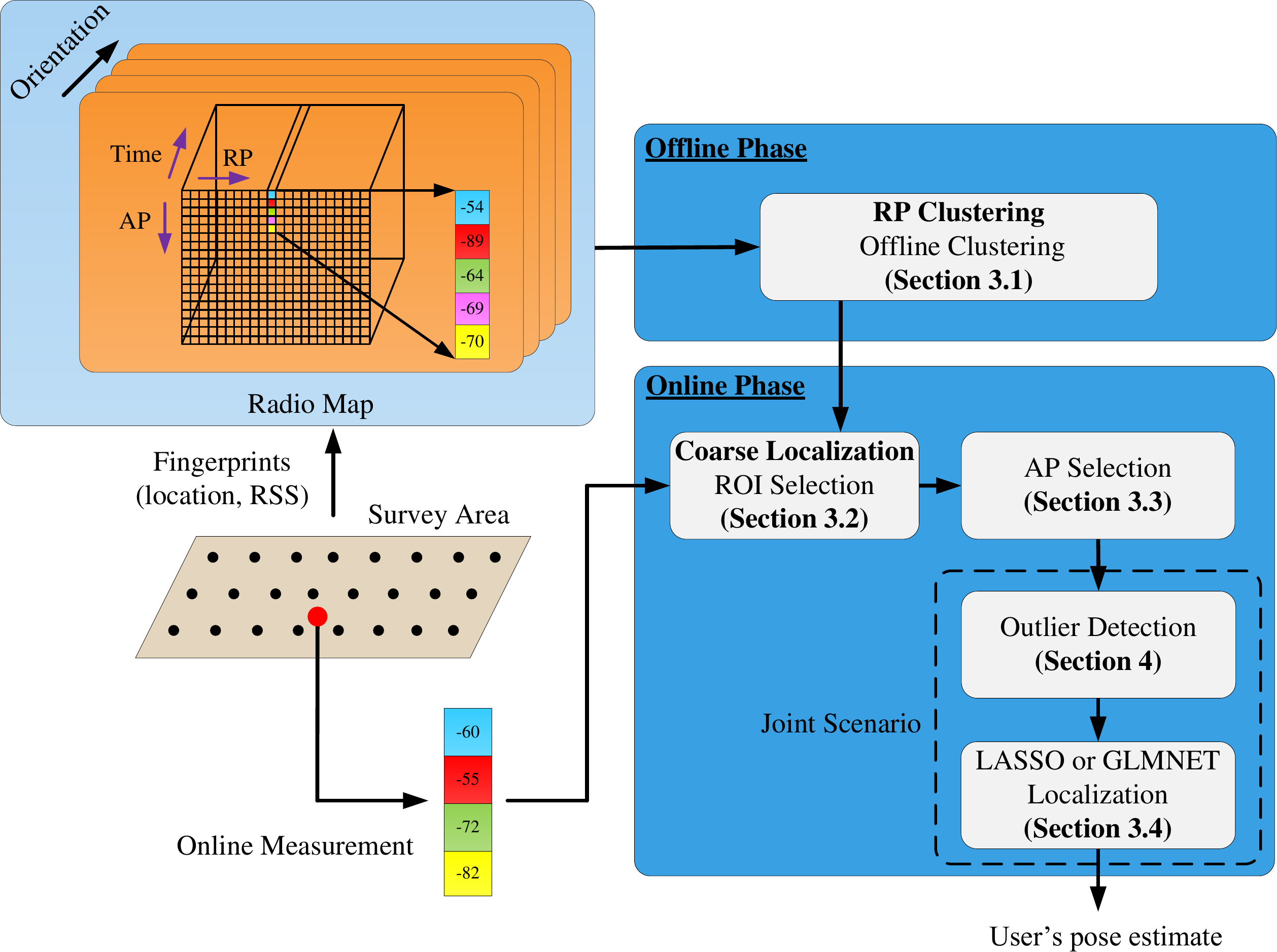}
           \centering

           \caption{Overview of the proposed indoor positioning scheme.}
           \label{figure1}
        \end{figure}

 The general framework of the proposed method is depicted in Fig. \ref{figure1}. We elaborate this framework in the order it appears in the diagram. 

\subsection{Offline Clustering}
The localization approach should be applicable on the handheld devices with low computational capabilities. Since the computational complexity of minimization problems increases polynomially with the size of the problem, searching for the mobile user's location in the entire RP set is impractical. Hence, the localization approach should confine the searching area into a subset of the RPs. On the other hand, the radio map is a database of fingerprints without any regularity in the order of the RPs. In this regard, the groups of RPs with similar features should be identified and the online measurement is compared against the representative features of each sub cluster. The clustering should be performed on the fingerprints in the offline phase so that the localization scheme deals with a categorized radio map. Since we record the radio map in four orientations, the offline clustering is applied at each orientation independently.
\par In our offline RP clustering approach, we consider each RP as the node of a graph. Let the set of all RPs be the nodes $\mathcal{V}$ of graph $\mathcal{G}$ with the edges $\mathcal{E}$ that connect the nodes: $\mathcal{G}=(\mathcal{V},\mathcal{E})$. A weight is assigned to each edge based on the \textit{similarity} of the nodes connected by the edge (the similarity will be defined shortly). The goal is to cluster this weighted graph so that the nodes with the most similarities are categorized in a cluster. For each cluster, we define a CH which represents the most features of that cluster. In our overlapped clustering scheme, a node might be a member or \textit{follower} (FL) of more than one cluster, and thus, is called a neighbor node. 
\par The weighted connection (edge) between two nodes is regulated by the similarity measure between the nodes. This similarity is based on the fact that spatially close RPs should receive similar readings from the same set of APs. The similarity metric that reveals this feature is the Hamming distance between the two corresponding RPs. Consider $\mathbf{p}_{j} \ \text{and} \ \mathbf{p}_{j'} $ are two RPs whose similarity needs to be determined. The hamming distance between these two nodes are defined as
\begin{equation}
\begin{split}
\mathbf{H}^{o}&(\mathbf{p}_{j} , \mathbf{p}_{j'})=d_{H}(\mathbf{I}_{j}^{o}, \mathbf{I}_{j'}^{o}  )= \sum _{i=1}^{L}|I_{j}^{i,o}-I_{j'}^{i,o}|\\
& \forall j,j'\in \{1,2,\ldots,N\}, j\neq j', \forall o\in \mathcal{O}
\end{split}
 \end{equation}
 where $ d_{H}$ is the Hamming distance between two \textit{reliability} indicator vectors $\mathbf{I}_{j}^{o}, \mathbf{I}_{j'}^{o}$ which specifically denote the set of reliable APs for $\mathbf{p}_{j} \ \text{and} \ \mathbf{p}_{j'} $ . An AP is considered reliable for $\mathbf{p}_{j} $ if its RSS fingerprints are above a threshold most of the time \cite{19} . To formalize this notion, first define the set of time slots that the surveyed radio map is above a threshold $\gamma$:
 \begin{equation}
 \label{eq6}
 \begin{split}
 \mathcal{T}_{j}^{i,o}&=\left\lbrace m \in \left\lbrace 1, \ldots , M\right\rbrace \arrowvert r_{j}^{i,o}(t_{m}) \ge \gamma \right\rbrace.\\
 \forall i&=1,\ldots,L, \ \ \forall j=1,\ldots,N, \ \ \forall o\in \mathcal{O}
 \end{split} 
   \end{equation} 
 The AP index $ I_{j}^{i,o}$, defined next,  denotes the APs whose readings satisfy \eqref{eq6} for 90\% of the time during recording the radio map fingerprints. This threshold can be experimentally set; the value 90\% is used in \cite{19}:
   \begin{equation}
   \label{eq7}
 I_{j}^{i,o}= \begin{cases}
          1 & \ \lvert \mathcal{T}_{j}^{i,o} \rvert \ge 0.9M\\
          0 & \text{otherwise}
      \end{cases} \ \ \ 
      i=1,\ldots,L, \ \ o\in \mathcal{O}
       \end{equation} 
    where $\left| \ . \ \right| $ denotes the cardinality. Next, let $\mathcal{L}_{j}^{o}$ be the set of APs that satisfy (\ref{eq7}):
        \begin{equation}
        \label{eq8}
      \mathcal{L}_{j}^{o}=\left\lbrace i \in \left\lbrace 1, \ldots ,L \right\rbrace  \big | I_{j}^{i,o}=1 \right\rbrace .
 \end{equation} 
\par  The similarity $s^{o}(j,j')$ between RPs $\mathbf{p}_{j}$ and $\mathbf{p}_{j'}$ is defined as
\begin{equation}
  \label{eq9}
  \begin{split}
  s^{o}(j,j')&=\begin{cases}
            \frac{1}{\mathbf{H}^{o}(\mathbf{p}_{j},\mathbf{p}_{j'})} &  \mathbf{H}^{o}(\mathbf{p}_{j},\mathbf{p}_{j'}) \ne 0\\
            \varLambda & \text{otherwise}
        \end{cases} \ \ \ 
   \\
  \forall j,j'&=1,\ldots, N \ , \ j\neq j' \ \ \ \forall o\in\mathcal{O}
  \end{split}
  \end{equation}
which is proportional to the inverse of Hamming distance between two different RPs, and $\varLambda$ is a sufficiently large number.  Since there are four orientations at each RP, the reliability of a typical AP is evaluated at each orientation individually. So, an AP may be reliable only in some orientations of an RP. The similarity measure renders the weight for each edge connecting the corresponding RPs in the weighted graph $\mathcal{G}$.
 \par In the following, the \textit{stability} of fingerprint readings is characterized by the variance of the measurements using repetitive measurements during the offline phase. The smaller variance in RP $j$ from AP $i$ indicates more stable measurements.  Hence, we compute the variance of readings for all RPs as
 \begin{equation}
 \begin{split}
  \Delta _{j}^{i,o}&=\frac{1}{M-1}\sum_{m=1}^{M} (r_{j}^{i,o}(t_{m})-\psi_{j}^{i,o})^{2} \\
 \forall i&=1,\ldots, L \ \ , \ \ \forall j=1,\ldots, N \ \ , \ \  \forall o\in\mathcal{O}.
 \end{split}
 \end{equation} 
 The variance of $\mathbf{p}_{j}$ is the average of variances in the set of APs that obey (\ref{eq7}), i.e. $\mathcal{L}_{j}^{o}$:
 \begin{equation}
 \begin{split}
 \Delta _{j}^{o}&=\frac{1}{\left| \mathcal{L}_{j}^{o} \right| }\sum_{i\in  \mathcal{L}_{j}^{o}}^{}\Delta _{j}^{i,o}\\
  \forall j=&1,\ldots, N \ \ , \ \  \forall o\in\mathcal{O}
 \end{split}
 \end{equation}
 The variance for all RPs is organized in a vector as 
 \begin{equation}
 \boldsymbol \Delta^{o}=[\Delta_{1}^{o}, \ldots, \Delta_{N}^{o}] \ \ \ \forall o\in\mathcal{O}.
 \end{equation}
 \par The pseudo-code of our clustering method is provided in Algorithm 1. The criteria for node $\mathbf{p}_{j'}$ to be in the cluster of $\text{CH}^{o}(k)$ is that the similarity between the two nodes should be greater than a predefined value as
\begin{equation}
\label{eq13}
\begin{split}
j'\in & \mathcal{\text{CH}}^{o}(k) \ \ \ \text{if} \ \ \ s^{o}(j',\text{CH}^{o}(k))\ge \eta  \\
&\forall k=1,\ldots,K \ \ \ \forall o\in\mathcal{O}.
\end{split}
\end{equation}
\begin{algorithm}[t!]
   \caption{: Offline Clustering}\label{a1}
   \begin{algorithmic}[1]
   \For{each direction $o \in \mathcal{O}$ }
   \For{all $j,j' \ \in \left\lbrace 1,\dots , N \right\rbrace $}
   \State compute  $ \mathbf{H}^{o}(\mathbf{p}_{j} , \mathbf{p}_{j'}) $
   \EndFor
   \For{all RPs  $\mathbf{p}_{j}$ }
   \State compute $\Delta_{j}^{o} $
   \EndFor
   \State CH candidates $\mathcal{B}^{o}=\left\lbrace 1,\ldots,N\right\rbrace $
   \State $k=0$;
   \While{$\mathcal{B}^{o} \ne \varnothing $}
    \State $k=k+1$
   \State select one node $j$ from $\mathcal{B}^{o}$
          \State define $\mathbf{p}_{j}$ as $\text{CH}^{o}(k)$
          \State $\mathcal{FL}^{o}(k)=\left\lbrace   j^{\prime} \ne j \rvert  s^{o}(j^{\prime},CH^{o}(k)) \ge \eta \right\rbrace $
          \State $\mathcal{C}^{o}(k)= \text{CH}^{o}(k) \cup \mathcal{FL}^{o}(k) $
          \State $\mathcal{B}^{o}=\mathcal{B}^{o} \backslash \mathcal{C}^{o}(k)$
     \EndWhile
      \State $K=\lvert\mathcal{C}^{o}(k)\rvert$
     \For {k=1,\dots,K}
     \State $\text{CH}^{o}(k)= \underset{j \in \mathcal{C}^{o}(k)}{\text{arg min} \ {\Delta_{j}^{o}}}$
     \EndFor
   \EndFor
   \end{algorithmic}
   \end{algorithm}
 where  $\mathcal{CH}^{o}=\left\{\text{CH}^{o}(1),\ldots, \text{CH}^{o}(K)\right\}$ is the set of all CHs for orientation $o\in\mathcal{O}$. Since the above criterion for each node may be satisfied for more than one cluster, each node has a table of CHs it belongs to. So, a node might be FL to more than one CH and is then considered as a boundary node between the two clusters that it belongs to. 
 \par Once all cluster members are defined, a representativity test is conducted within each cluster to select the best node as the representative node of that cluster, CH (see lines 19-21 in Algorithm 1). This may lead to switching the CH of that cluster. This test measures the suitability of the CH to represent the characteristics of its followers.
 The cluster, $\mathcal{C}^{o}(k)$, is the set comprising of the cluster head, $CH^{o}(k)$ and its followers, $\mathcal{FL}^{o}(k)$ 
 \begin{equation}
 \mathcal{C}^{o}(k)=\left\lbrace  \text{CH}^{o}(k) \right\rbrace \cup \mathcal{FL}^{o}(k).
 \end{equation}
  A node is selected as the CH when it has the least variance of fingerprints amongst all the cluster members
\begin{equation}
\begin{split}
\text{CH}^{o}(k)&=\left\lbrace \mathbf{p}_{j} | \Delta_{j}^{o}=\text{min} \ \left\lbrace \Delta_{l}^{o}\right\rbrace , l=1,\ldots, \lvert\mathcal{C}^{o}(k) \rvert \right\rbrace \\
\forall k&=1,\ldots,K \ \ \ \ \ \forall o\in\mathcal{O}.
\end{split}
\end{equation}
\par The offline clustering defines a localization metric which evaluates the similarity between a FL with its CH and possible CH replacements based on previously defined stability measure.

\subsection{Online Region of Interest (ROI) Selection}
 Since solving the optimization algorithms are the most computationally intensive part, the size of the problem should also be reduced. Hence, we find an ROI as the most probable region on which the user exists. Selecting ROI is based on a comparison between the reliability vectors of the online measurements and the radio map. After the user reads the online measurement $ \boldsymbol y $, an indicator vector $\mathbf{I}_{\boldsymbol y}=\left\lbrace I_{\boldsymbol y}^{1},\dots, I_{\boldsymbol y}^{L} \right\rbrace $ is specifically defined to denote the set of reliable APs. An AP is considered reliable if its online RSS values is above a predetermined threshold
\begin{equation}
\label{eq16}
I_{\boldsymbol y}^{i}=
\begin{cases}
         1 & \text{if } y^{i}\ge \gamma\\\
         0              & o.w.
     \end{cases}.
\end{equation}
This reliability vector defines the most trustable APs in online measurements that we can engage in computations. Similarly, we have defined the AP reliability matrix for the radio map in Section 3.1. For comparing the online measurement vector, $\boldsymbol y$, with radio map, $\boldsymbol \Psi^{o}$, the Hamming distance between the online measurement \text{reliability} and CH \text{reliability} is defined as
 \begin{equation}
 \begin{split}
 \mathbf{H}^{o}_{\text{CH}(k)}( \mathbf{I}_{\boldsymbol y},  \mathbf{I}_{\text{CH}(k)}^{o})=\sum_{i=1}^{L}\left| I_{\boldsymbol y}^{i}-I_{\text{CH}(k)}^{i,o}\right| \\
 k=1,\ldots,K \ , \ \ o\in \mathcal{O}.
 \end{split}
 \end{equation}
 At each orientation, the CH with the minimum distance is chosen and the corresponding cluster is included in the modified radio map, $ \tilde{\mathbf{\Psi}} $. If the cluster with the minimum distance has points common with other clusters, then the neighbor cluster RPs are also included. The modified radio map, $ \tilde{\mathbf{\Psi}} $, may contain the fingerprints for one RP but in different orientations. The set of selected RPs in the ROI is denoted as
 \begin{equation}
 \mathcal{\tilde{P}}\subset \mathcal{P}, \lvert \mathcal{\tilde{P}} \rvert =\tilde{N}\le N
 \end{equation}
 The Pseudo-code of the ROI selection method is provided in Algorithm \ref{a2}. We use MATLAB notation for matrix concatenation.
 
 \begin{algorithm}[t!]
    \caption{: ROI Selection}\label{a2}
    \begin{algorithmic}[1]
    \State compute $\mathbf{I}_{\boldsymbol y}$
    \State $\tilde{\mathbf{\Psi}}=[]$
    \For{each orientation $o \in \mathcal{O}$ }
    \State compute $\mathbf{H}^{o}_{CH(k)}$ for $k=1,\dots,K$
    \State find $k' \ \text{s.t.} \  \mathbf{H}^{o}_{CH(k')}=\underset{k}{\text{min}}\left\lbrace \mathbf{H}^{o}_{CH(k)} \right\rbrace $
    \State $\mathcal{K}'=\left\lbrace k=1,\ldots,K \big|\mathcal{C}^{o}(k')\cap \mathcal{C}^{o}(k) \ne \varnothing\right\rbrace  $ 
    \State$\tilde{\boldsymbol \Psi}^{o}=\left\lbrace \boldsymbol \psi_{j}^{o}\big | j \in \mathcal{C}^{o}(k') \ \text{and} \ j \in \mathcal{C}^{o}(k) \ \ \forall k \in \mathcal{K}' \right\rbrace $
    \State$\tilde{\mathbf{\Psi}}=\left[\tilde{\mathbf{\Psi}}, \tilde{\mathbf{\Psi}}^{o}\right]$
    \EndFor
    \end{algorithmic}
    \end{algorithm}
 \subsection{AP Selection}
 In principle, localization in a two dimensional space needs three anchors based on which the ambiguities are removed. The received signal strength in the user's device is a function of the distance between the AP and the user's location. However, in indoor settings, the intermediate environmental factors such as shadowing and multipath highly affects the received signal at the user's end. So, the monotonic relationship between the received signal strength and the user-AP distance is not preserved. This nonlinear behavior causes the user to receive similar measurements from distinct APs and increases the bias in the estimates. To alleviate this effect, positioning schemes should select a suitable subset of available APs. Several procedures have been studied in the literature. The strongest AP selection method selects a subset of the APs with the strongest RSS measurements, assuming that stronger signals are more reliable. Also, the well known Bhattacharyya distance measures the distance between the probability densities of the radio map fingerprints of two APs \cite{46}. For simplification, the assumption is that the RSS distributions are Gaussian. Another approach relies on the information potential, which measures the distance between two probability density functions using the Gaussian kernel \cite{47}. However, this method needs an exhaustive search over a large set whose size is proportional to the area and number of available APs. In addition, the Gaussian assumption is not necessarily valid in practical setup and has been reported to be violated \cite{48}. 
  \par Vendors have to setup several APs to provide the wireless network over the whole area due to the range limitation of the APs. Most of the localization schemes select the APs based on the fingerprints in offline phase blindly to the location of the user. Nonetheless, a fixed subset of APs cannot represent the features of the whole environment especially for large areas such as indoor offices, malls, and airports. The APs selection scheme should examine the suitability of the APs tailored to the location of the user.  The Fisher criterion can be employed to quantify the discrimination ability of an AP over the RPs \cite{19}
  \begin{equation}
  \label{fisher}
  \begin{split}
 \zeta^{i,o}&=\frac{\sum_{j\in \mathcal{\tilde{P}}}^{}(\bm \psi_{j}^{i,o}-\bar{\bm \psi}^{i,o})^{2}}{\frac{1}{M-1}\sum_{m=1}^{M}\sum_{j\in \mathcal{\tilde{P}}}^{}(\mathbf{R}_{j}^{i,o}(t_{m})-\bm \psi_{j}^{i,o})^2}
  \end{split}
  \end{equation}
  where $ \bar{\bm \psi}^{i,o}= \frac{1}{\lvert\mathcal{\tilde{P}}\rvert}\sum_{j\in \mathcal{\tilde{P}}}^{} {\bm \psi}_{j}^{i,o} $ (see e.g. \cite{47} for a general definition of the Fisher criterion). The numerator defines how fingerprints are spread over the RPs and the denominator captures the variability of fingerprints. So, this criterion indicates the differentiability between the RPs and assigns low scores to unstable APs. An AP score is obtained through averaging the Fisher scores in all four orientations. 
  \par The Fisher criterion selects the APs based on their statistical properties  and has been previously used as an AP selection method in the offline phase. However, in our method, it is used in the online phase as only the fingerprints that are in ROI are selected to participate in \eqref{fisher}. In this way, the subset of APs that are more suitable for the user is selected.
  \begin{algorithm}[t!]
      \caption{: AP Selection}\label{a3}
      \begin{algorithmic}[1]
      \For{each orientation $o \in \mathcal{O}$ }
      \State compute $ \zeta^{i,o}$ for $i=1,\ldots,L \ \ , \ \ j\in \mathcal{\tilde{P}}$
      \EndFor
      \State $\varsigma^{i}=\frac{\sum_{o \in \mathcal{O}}^{}\zeta^{i,o}}{\lvert\mathcal{O}\rvert}$
      \State Define $\tilde{\mathcal{L}}$ as the set of APs with the $\tilde{L}$  largest Fisher scores $\varsigma^{i}$, for a predetermined  $\tilde{L}= \lvert \tilde{\mathcal{L}}\rvert $.
      \end{algorithmic}
      \end{algorithm}
 \par The pseudo-code for AP selection method is provided in Algorithm \ref{a3}. The proposed method does not depend on probability density function models and has thus favorable computational complexity. Let $ \tilde{\mathcal{L}} \subset \mathcal{A} $, $ \lvert  \tilde{\mathcal{L}}  \rvert \le L $, be the set of $ \lvert  \tilde{\mathcal{L}} \rvert $ APs with the greatest Fisher score. The matrix $ \mathbf{\Phi} $ is the AP selection matrix whose $i$-th row, $ \mathbf{\Phi}^{i} $, is a $ 1\times L $ vector that defines the selected APs through zeroing out all the indices except the selected AP index as
  \begin{equation}
  \mathbf{ \Phi}^{i}=[\ldots,\underbrace{1}_\text{Index of selected AP},\ldots] \ \ \ \ \ \forall i=1,\ldots, \lvert \tilde{\mathcal{L}} \rvert.
  \end{equation}
  
   \subsection{Localization Scheme}    
  \par The typical problem in WLAN positioning is to estimate the user's location using fingerprinting data on a discrete grid of RPs whenever the user reads a RSS observation in online phase. Basically, we look for a function that maps the radio map in conjunction with the online measurements to a unique subset of RPs: $  \hat{\boldsymbol \theta}=f(\mathbf{R},\boldsymbol y) $. Although sparse recovery minimization (\ref{eq4}) renders sparse vectors, the reconstructed position vector is likely to contain erroneous nonzero indices counting on less probable RPs. Also, the similarity between the online measurements and radio map is of great importance. So, the residuals $ \mathbf{r}=\mathbf{y}-\boldsymbol \Phi  \tilde{\boldsymbol \Psi} \boldsymbol \theta$ should also be small. Considering the conventional CS WLAN positioning, the problem can be modified if the optimization algorithm also suppresses the error between the online response variable and predictor vectors. Hence, we propose sparse recovery algorithms that include the $\ell_{1} $-norm of the location vector and $\ell_{2} $-norm of the residuals simultaneously. The positioning convex optimization problem can be reformulated as
  \begin{equation}
  \label{eq17}
 \hat{\boldsymbol \theta} =\underset{\boldsymbol \theta}{\text{argmin}} \ \left[ \frac{1}{ \lvert \tilde{\mathcal{L}} \rvert}\Arrowvert \mathbf{y-\mathbf{H} \boldsymbol \theta}\Arrowvert_{2}^{2} + \lambda \Arrowvert\boldsymbol \theta\Arrowvert_{1}\right]
  \end{equation}
  where $ \mathbf{H}=\mathbf{\Phi\tilde{\Psi} }$ and $ \lambda \ge 0$ is a tuning parameter. This problem is also known as $\ell_{1} $-penalized least squares, LASSO, which incorporates feature and model selection into the optimization \cite{63}. The parameter $ \lambda$ is a regularization parameter that needs to be adjusted properly. The first component looks for coefficients that minimize the residuals, and the position vector become a sparse vector through \textit{soft thresholding} on the second term. The LASSO uses only the RPs with the most similar features thus simplifying the model automatically. As stated before, once the radio map fingerprints are recorded in four orientations, the coarse localization method renders a modified radio map whose columns may include readings from the same AP in different orientations. If the propagation environment is similar in different orientations, the fingerprints in the modified radio map, $ \tilde{\mathbf{\Psi}} $, may have similarities, leading to correlated predictors. LASSO has shown to be more indifferent to these correlated predictors.
  \par Consider there are correlated predictors in the modified radio map. If the user is exactly in an RP, the online measurement is supposed to be very similar to the fingerprints of that RP and the modified radio map may contain some of the orientations of that point. Another possible case is when the user is between two RPs with similar environmental features. In both cases the location estimation problem is expected to assign higher coefficients to the points with correlated fingerprints. Hence, the correlated predictors should be allowed to jointly borrow strength from each other. We propose another convex optimization problem which incorporates the above features:
  \begin{equation}
  \label{eq18}
  \begin{split}
\hat{\boldsymbol \theta} &=\underset{\boldsymbol \theta}{\text{argmin}} \ \left[ \frac{1}{\lvert  \tilde{\mathcal{L}} \rvert}\Arrowvert \mathbf{y-\mathbf{H }\boldsymbol \theta}\Arrowvert_{2}^{2} +  P_{\alpha}  \right]\\
P_{\alpha}&=\lambda \bigl((1-\alpha)\Arrowvert \boldsymbol \theta \Arrowvert _{2}^{2}+\alpha \Arrowvert \boldsymbol \theta \Arrowvert_{1})\bigr)
  \end{split}
  \end{equation}
  where $ \lambda \ge 0$ is a complexity parameter and $0\le\alpha\le1 $ is a compromise between \textit{ridge regression} and LASSO. Ridge regression promoted to the shrinkage of the coefficients of correlated radio map columns towards each other and is expressed by the $\Arrowvert \boldsymbol \theta \Arrowvert _{2}^{2}$ objective. Hence, we take advantage of the correlation between the radio map readings. If $\alpha=0 $, the above formulation amounts to the \textit{Ridge Regression}. As  $\alpha$ increases from $0$ to $1$ for a given $\lambda$ the sparsity of the solution increases monotonically from $0$ to the sparsity of the LASSO solution. So, this problem jointly considers the correlated predictors and finds a sparse solution for the user's pose. This problem is known as GLMNET \cite{64}. 
  \par The proposed sparse recovery formulations have several advantages over (\ref{eq4}). First, CS works for noiseless measurements, which is not practical for WLAN fingerprinting localization as the measurements contain errors. Also, the CS relies on orthogonality properties of $\mathbf{H}$ to minimize any correlation between the RSS readings while basically utilize correlations to find a more suitable sparse solution.
  \par The computational complexity of the above optimization problem grows with the number of predictors (size of radio map) and dimension of vector $\boldsymbol \theta$. Therefore, the previously mentioned coarse localization scheme reduces the size of the area that the optimization problems needs to seek for the solution and hence reduces the computational time for solving the problem. This allows these procedures to be executed on resource-limited devices.   
  \par If online measurements differ substantially from the radio map fingerprints, as the radio map becomes less reliable through the time, or the mobile user is not located exactly at one RP, the solution to (\ref{eq17}) or (\ref{eq18}) may not be exactly a 1-sparse vector and may contain several significant nonzero coefficients. Hence, a post-processing step after obtaining the estimated coefficients is needed. We choose the dominant coefficients in $\hat{\boldsymbol \theta}$ that are greater than a threshold $ \beta $, and take the weighted average of the corresponding RP positions  . Let $\mathcal{J}_{\beta}$ denote the indices of the $ \hat{\boldsymbol \theta}$ whose values are above a certain threshold $ \beta $ as:
  \begin{equation}
  \label{eq23}
   \mathcal{J}_{\beta} =\{v \in \left\lbrace 1,\dots, V \right\rbrace |\hat{\boldsymbol \theta} (v) \ge \beta\}
  \end{equation}
  where $V$ is the number of columns of $\tilde{\bm \Psi}$. Let $\tilde{\mathcal{P}}_{\beta}$ be the set of RPs $\tilde{\mathbf{p}}_{v}$ corresponding to $\hat{\boldsymbol \theta} (v)\ge \beta$ and allow repetition of elements in $\tilde{\mathcal{P}}_{\beta}$ if there are more than one element in $\mathcal{J}_{\beta}$ corresponding to $\tilde{\mathbf{p}}_{v}$. The location of the mobile user is computed as the centroid of the convex hull generated by the RPs in $ \tilde{\mathcal{P}}_{\beta}$:
  \begin{equation}
  \hat{\mathbf{p}}=(\hat{x},\hat{y})=\frac{1}{\sum_{v \in \mathcal{J}_{\beta} }^{}\hat{\boldsymbol \theta}(v)} \sum_{v \in \mathcal{J}_{\beta}}^{}{\hat{\boldsymbol \theta}}(v)\cdot \tilde{\mathbf{p}}_{v}
  \end{equation}
  where $\tilde{\mathbf{p}}_{v}=(\tilde{x}_{v},\tilde{y}_{v})\in \tilde{\mathcal{P}}_{\beta}$. 
   \begin{figure*}[t!]
                    \includegraphics[scale=0.945,angle =90]{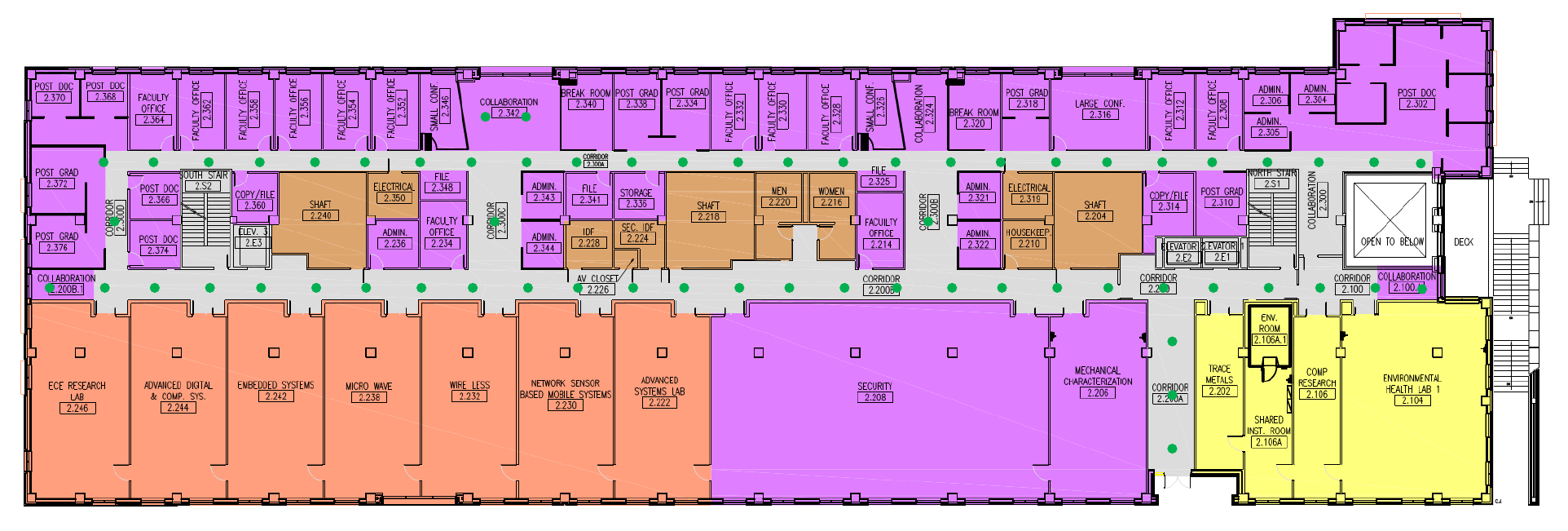}
                         \centering
                         \caption{The actual map of the experimental environment. Green dots denote the reference points.}
                         \label{figure3}
                      \end{figure*}
                      
  \section{Outlier Detection}
In the previous section, we elaborated on the proposed sparsity based indoor positioning approach. However, despite the computational efficiency of sparsity-based approaches, indoor WLAN fingerprinting-based positioning systems still face challenges. The WLAN fingerprints can be easily affected by the shadowing and multipath effects in complex indoor environments that impair the estimation accuracy. APs' fingerprints are also prone to other problems as the WLAN infrastructure is not essentially established for positioning purposes. Modern APs adapt their transmitting power according to the serving traffic. If the traffic load during the online positioning is different than that of fingerprinting, the APs readings may be quite different. In addition, APs are vulnerable to adversary attacks which may impersonate, jam, or corrupt the data by attenuating and amplifying the APs RSS readings. Our real data measurements also indicate that APs' readings may not be available due to transient effects in either APs or the receiver's Network Interfaces (NIs). Likewise, APs that were available during fingerprinting may not be available in the online phase, and vice-versa. These previously mentioned phenomena seriously challenge the assumptions in the literature and make them less practical. All these deficiencies are modeled as outliers. This section develops algorithms for positioning in the presence of outliers.
\par Outliers may occur in both the offline and online phases, however, we assume that preventative measures, including fingerprinting over a long period as well as validation and attack detection have taken place on the offline RSS fingerprints \cite{53}. Hence, our main focus is outlier detection in the online phase.
\par An outlier occurs when the online measurement from an AP is significantly different from any fingerprint in the area. The existing AP selection schemes select the APs based on the AP performance during the fingerprinting period. However, if an AP provides highly different readings in online positioning phase, the online measurements and the radio map fingerprints may have a large deviation. This phenomenon has surprisingly received little attention in the existing literature, which motivates us to adapt the schemes of Sec. 2 and Sec. 3 so that they can detect the outliers in the online measurements. We show how the concept of sparsity-promoting regression can be utilized towards joint outlier detection and positioning.  
\par The outlier detection scheme in this paper relies on explicit modeling of the outliers present in the online measurements. Specifically, with $\bm \kappa$ denoting the outlier vector, the online measurements adhere to the following model which extends (\ref{eq2}):
\begin{equation}
\label{eq24}
\mathbf{y}=\mathbf{\Phi \Psi \boldsymbol\theta +\boldsymbol \kappa+\boldsymbol\epsilon}.
\end{equation}
The advantage of the previous model is that the outlier vector $\boldsymbol \kappa$ will be sparse as long as the number of corrupted APs is small, and can therefore be estimated jointly with the position indicator vector $\boldsymbol \theta$ via $\ell_{1}$-minimization. The premise of explicitly modeling the outliers for robust regression in general statistical settings has been previously analyzed in \cite{50} and \cite{51}. In what follows, the CS, LASSO, and GLMNET approaches are modified so that the outlier vector $\bm\kappa$ can be estimated alongside the user's position vector $\bm\theta$.
\par The modified CS (M-CS) approach minimizes the weighted combination of the $\ell_{1}$ norms of $\bm\theta$ and $\bm\kappa$.
\begin{equation}
\label{eq26}
\begin{split}
 (\hat{\boldsymbol\theta}, \hat{\boldsymbol\kappa})&=\underset{\boldsymbol\theta,\boldsymbol \kappa }{\text{argmin}} \ \left[ \Arrowvert \boldsymbol\theta \Arrowvert_{1} +\mu \Arrowvert \boldsymbol \kappa \Arrowvert_{1} \right] \\
  & s.t.\  \mathbf{y}=\mathbf{\Phi \Psi \boldsymbol\theta+\boldsymbol \kappa}
\end{split}
\end{equation}
\par The modified LASSO (M-LASSO) approach minimizes the squared residuals, in addition to the  $\ell_{1}$  norms of the sparse vectors:
  \begin{equation}
  \label{eq27}
 (\hat{\boldsymbol\theta}, \hat{\boldsymbol\kappa}) =\underset{\boldsymbol\theta,\boldsymbol \kappa }{\text{argmin}} \ \left[ \frac{1}{ \lvert \tilde{\mathcal{L}} \rvert}\Arrowvert \mathbf{y-\mathbf{H} \boldsymbol \theta}-\boldsymbol \kappa\Arrowvert_{2}^{2} + \lambda \Arrowvert\boldsymbol \theta\Arrowvert_{1}+\mu \Arrowvert \boldsymbol \kappa \Arrowvert _{1}\right]
  \end{equation}
  where $ \mu $ is a tuning parameter. 
  \par Finally, the modified GLMNET (M-GLMNET) amounts to the following optimization problem:
  \begin{equation}
  \label{eq28}
  \begin{split}
  (\hat{\boldsymbol\theta}, \hat{\boldsymbol\kappa}) =\underset{\boldsymbol\theta,\boldsymbol \kappa }{\text{argmin}} \ \left[ \frac{1}{\lvert \tilde{\mathcal{L}} \rvert}\Arrowvert \mathbf{y-\mathbf{H} \boldsymbol \theta -\boldsymbol \kappa}\Arrowvert_{2}^{2} + P_{\alpha} \right]\\
   P_{\alpha}=\lambda \bigl((1-\alpha)\Arrowvert \boldsymbol \theta \Arrowvert _{2}^{2}+\alpha \Arrowvert \boldsymbol \theta \Arrowvert_{1})+\mu \Arrowvert \boldsymbol \kappa \Arrowvert _{1}\bigr).
  \end{split}
  \end{equation}
  \par In the previous joint localization and outlier detection formulations, the outlier vector, $\boldsymbol \kappa$, enables the optimization algorithm to discard the outliers in the online measurement vector. Both user's location and the outlier indicator vector have the weights $ \lambda $ and $ \mu $, respectively.
  
 \section{Implementation Results and Discussion}
   This section elaborates on the experimental evaluation of the proposed methods on a real environment. The mobile recording device and environment setup are discussed, and the results are analyzed.
   \subsection{RSSI Capturing Device}
   Since a large number of mobile handheld devices are functioning on Linux-based Android operating system, the RSS fingerprints have been collected using a self-developed Android user application on a Samsung tablet (Galaxy Tab A) functioning on Android Lollipop 5.0.2 using the inherent android.net.wifi package. The wifi package gives the opportunity to manually set the reading interval of the device Network Interface Card (NIC); however, our experiment showed that the readings are repeated for small sampling times. A possible reason might be that the device has an inherent delay in refreshing the  NIC buffer. For this reason, we set the device to record the RSSI readings once every second and store the MAC address and RSS in a database.
   \subsection{Environment Setup}
   Real data were obtained from the third floor of the five-story Applied Engineering and Technology (AET) building at the University of Texas at San Antonio (UTSA). The dimensions of the surveying site are 300 ft $\times$ 35 ft, which is comparable to those reported in \cite{19,r27}. The layout of the experimentation area is shown in Fig. \ref{figure3} and the green circles show the fingerprinting locations. The area is representative of an indoor office environment with a complex wireless propagation pattern due to research labs, offices, library, and study areas and has a high volume of commuting. For fingerprinting, the RP locations can be drawn from a point process, where the position of the RPs come from a probability distribution. This method leads to obtaining points along the most traveled routes. Another approach is to consider a predefined uniform grid of RPs. A uniform grid RPs has been utilized by most of the previous works in the literature and facilitates the comparisons with the existing methods. Due to the presence of the obstacles such as walls, furnitures, and lab instruments, the RPs formed an almost uniform grid in our experimental setup (Fig. \ref{figure3}).
   \par  Another important feature is the distance between the RPs, known as granularity. As the Wifi signals are not originally designed for localization, they do not contain overheads that differentiate amongst regions in an area. So, a dense granularity of the grid points impose a great redundancy in the localization system with a high computational cost. In other words, collecting dense fingerprints may not necessarily lead to a finer localization precision. On the other hand, a denser grid could be beneficial in sparsity-based methods, because the measured fingerprint would be more likely to be produced by a 1-sparse position vector. Existing methods generally design their settings with 5 ft and 9 ft spacing. In our work, we have collected the fingerprints on a 3 ft grid spacing.
  
   \subsection{Training and Testing Data Collection}
   \par During the offline phase, the area has been divided into a total of 192 grid points and we collected 100 samples per grid point at four device orientations over three weeks. The fingerprinting time has been chosen during the office hours to capture the most nonlinear RSS features when the area was experiencing a high volume of commuters. A total of 268 different MAC addresses have been visible, although, the number of APs is not known exactly. Hence, we assume that each MAC address is associated with one AP.
   \par The testing measurements have been captured on and off the grid points. A collection of $N_{t}=100$ random points is selected in the surveying area and only a single RSS measurement is read at each point. Finally, the tuning parameters for \eqref{eq13}, \eqref{eq6} and \eqref{eq16} are listed in Table 1.
      \begin{table}[t]
      \label{table1}
      \begin{center}
       \caption{Tunning Parameters Setting for the Proposed Schemes}
      \begin{tabular}{ c || c }
       \hline
      $\gamma$  & -70 dBm\\ \hline
      $\eta$  & 0.92$L$ \\ \hline
       
      \end{tabular}
       \end{center}
      \end{table}
 
   \subsection{Localization Error}
   The performance of the proposed methods is measured by computing the average Euclidean distance between the estimated and the true locations of the random points, known as Mean Average Error (MAE)
      \begin{equation}
      \text{MAE}=\frac{1}{N_{t}}\sum_{j=1}^{N_{t}} \sqrt{ \left( \hat{\mathbf{p}}(j)-\mathbf{p}(j)\right) ^{T}\left( \hat{\mathbf{p}}(j)-\mathbf{p}(j)\right) } 
      \end{equation}
   where $\mathbf{p}(j)$ and $\hat{\mathbf{p}}(j)$ are the true and estimated locations, respectively. The Localization error of a positioning system is highly affected by RSS variations, the number of APs, and the subset of the RPs and APs chosen in the localization system. For alleviating the time-varying RSS fingerprints, the recorded offline readings were time-averaged as explained in Sec. 2. The online measurement is captured once. The assumption of capturing several online measurements is unrealistic in existing localization methods unless the user is staying at that location during the online capturing time. Hence, in our approach, only a single online measurement is used for localization.
   \par The computation efficiency of the proposed methods are also compared with other approaches by evaluating the average running time of online location estimation for all test points divided by their number. The running times are reported for an Intel (R) core (TM) i5 with 3.2 GHz CPU.
   \par In the following sections, we illustrate the effects of the clustering (ROI selection), AP selection, outlier detection, and computational efficiency of the proposed localization scheme.
   \begin{figure}[t!]
         \includegraphics[scale=0.545]{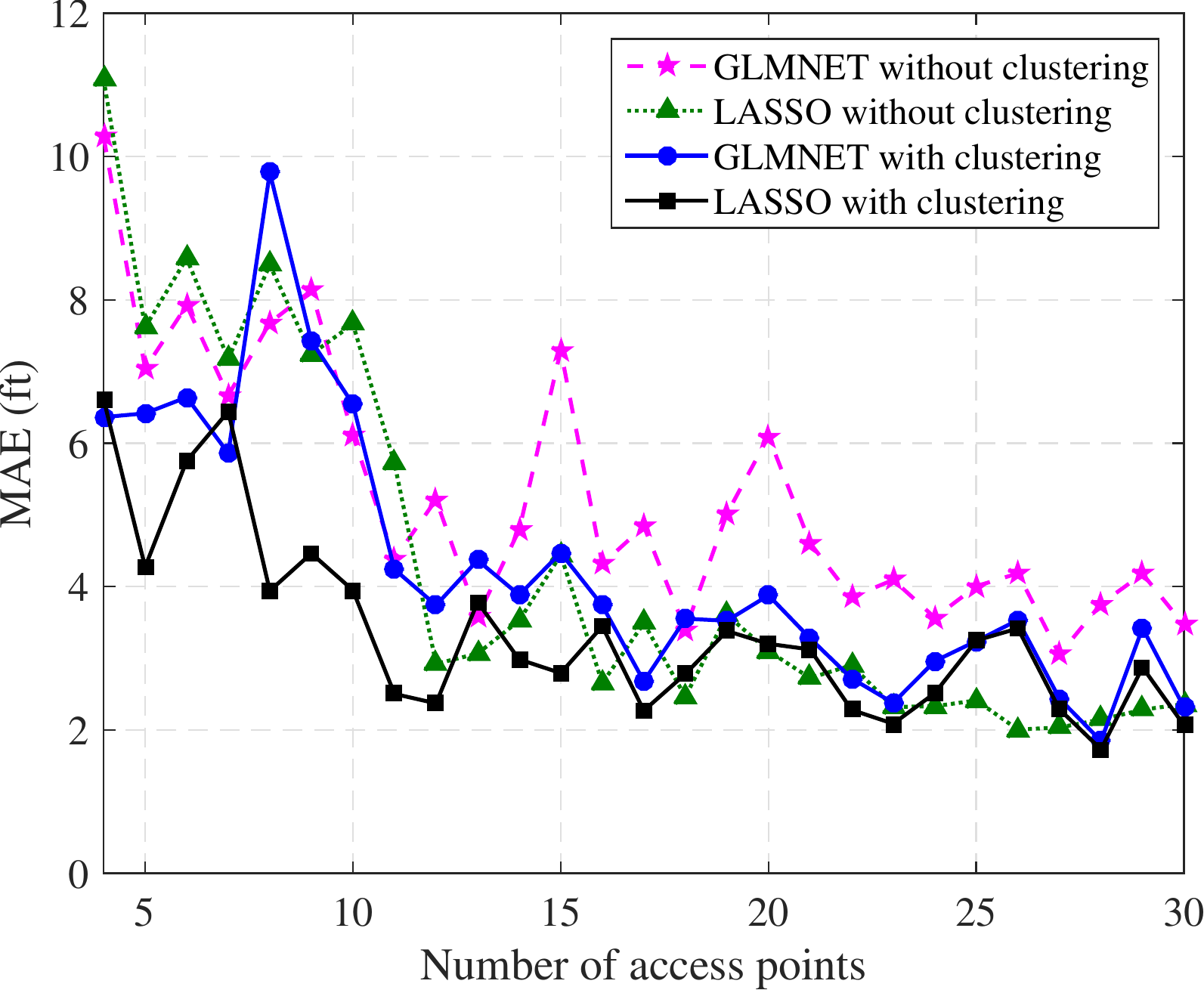}
            \centering      \caption{The MAE for LASSO and GLMENT localization with respect to different number of access points, with and without offline RP clustering.}
                              \label{figure4}
                           \end{figure}
                           
              \begin{figure}[t!]
                       \includegraphics[scale=0.51]{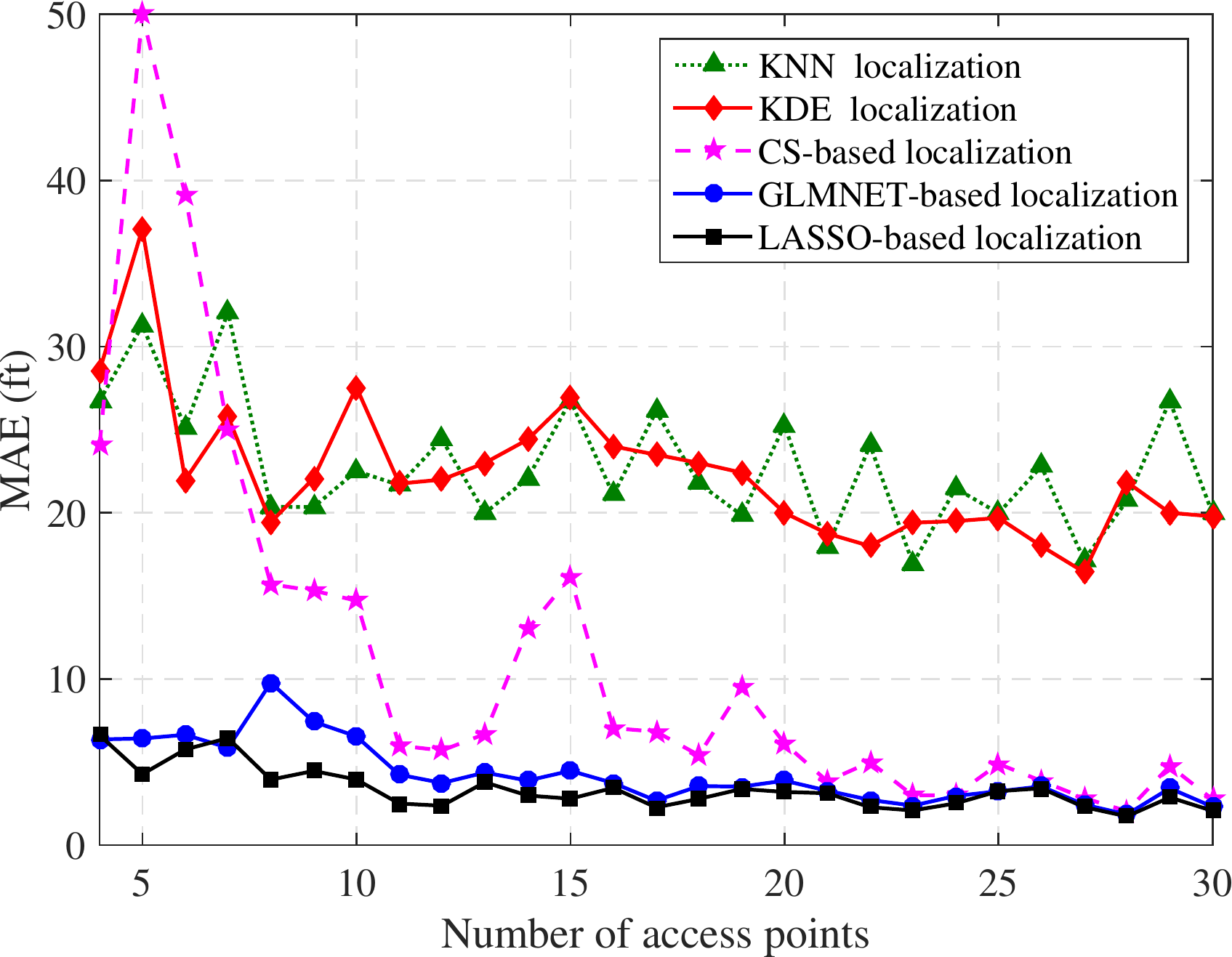}
                            \centering
                            \caption{The average localization error for KNN, KDE, CSS-based, Lasso-based, and GLMNET-based localization for various number of access points without outliers.}
                            \label{figure5}
                         \end{figure}
         
   \subsection{Offline Clustering}
   The performance of the sparse recovery localization approaches highly depends on the number of online measurements (APs) and the number of predictors (RPs). In CS-based localization, the number of measurements should be in the order of $\text{log }\tilde{N}$ to satisfy a unique recovery of the position vector. However, if the radio map readings are correlated, the required number of measurements greatly increases and hence more APs should be used for a good sparse recovery. This problem is alleviated in LASSO and GLMNET localization as each has its own recovery scheme for correlated predictors. Clustering brings a two-fold benefit for the localization. Since the solution should be found on the number of predictors (RPs), clustering helps to decrease the number of RPs, the columns of $\tilde{\boldsymbol \Psi}$, and hence, the number of APs needed for a good recovery decreases. Clustering also decreases the size of the problem which leads to a lower computational effort that is crucial for handheld devices. To this end, we have studied the localization error of the GLMNET and LASSO localization schemes. Fig. \ref{figure4} shows the MAE of the proposed localization schemes for an increasing number of APs. The difference between the errors for a lower number of APs is clear and shows that the clustering has a great effect on the LASSO and GLMNET localization error. For instance, 8 APs are sufficient to reduce the position error to nearly 4 ft. When 12 or more APs are used, the position errors of the LASSO approach with and without clustering are similar. So, the clustering helps the localization when a small subset of APs are used.

\subsection{Online Localization and Comparison to  Prior Work}
The first stage of the online localization is the ROI selection. After the most probable subset of the area has been defined by the ROI selection stage, the mobile user's position is estimated through the fine localization scheme. The fine location estimation is dependent on the number of APs used for the positioning. We have investigated the localization error with different number of APs for the two proposed localization methods and compared the result with the Weighted KNN (WKNN) \cite{101}, Kernel Density Estimation (KDE)  \cite{19}, Contour-based \cite{99}, and CS-based localization method \cite{r27}. These methods have also been implemented in our setting. For WKNN, the 10 most similar RPs with their corresponding weights have been used (K=10). For Contour-based localization, the AP transmit power and the path loss coefficients are estimated similarly to EZ \cite{60}, using nonlinear least squares. Our duplicated version of CS-based localization approach contains all the proposed steps in \cite{r27}, except for the clustering and coarse localization schemes, which have been substituted with our proposed offline clustering of Sec. 3.1 and ROI selection of Sec. 3.2, respectively. 
\par  Fig. \ref{figure5} compares the MAE for the KNN, KDE, and CS-based localization with the proposed schemes versus the number of APs. In this set of results, we assume that the APs contain no outliers. The KNN and KDE approaches give by far less accurate location estimates and render a localization error for more than 20 ft  when less than 20 APs are used. However, the results for the proposed approaches illustrate that increasing the number of APs from 4 to 12 reduces the positioning error by half. The localization accuracy is slightly enhanced thereafter when more APs are used for positioning. It is concluded that 5 APs give an acceptable position error while increasing small computational effort. However, the  CS-based localization introduces large errors when the number of APs is low while the estimation error decreases abruptly if more than 12 APs are used. The CS-based localization always introduces larger localization error even if more APs are used for localization. We observed that increasing the number of APs does not necessarily decreases the localization error of the CS-based method as this method does not consider the correlated fingerprints. LASSO shows a better performance when the number of APs is small; however, both GLMNET and LASSO give the same performance with a large number of APs. 
 \begin{table*}
                      \label{compare}
                     \begin{center}
                     \caption{Position Error Statistics and Running Times of the Developed Method Compared with Implemented Competing Approaches}
                     \begin{tabular}{ |c|c|c|c|c|c|c| } 
                     \hline
                     Methods  & 25\% (ft)& 50\% (ft)& 75\% (ft) & 100\% (ft) & Comments & Time (ms)\\
                     \hline
                     WKNN \cite{101} & 6.69 & 12.86 & 27.34 & 190.62&  K=10& 0.138 \\ 
                       \hline
                     KDE \cite{19}& 3.84 & 11.71 & 28.75 & 217.85 &  Gaussian Kernel is used& 134 \\ 
                       \hline
                     Contour-based \cite{99}& 56.84 & 125.31 & 187.25 & 261.5 & Estimates path loss parameters  & 312\\ 
                       \hline
                       CS \cite{r27}& 1.87 & 4.48 & 13.64 & 335.91& Solved via l1-magic \cite{103} & 2.21\\
                                              \hline
                     GLMNET&  1.3  & 3.16 & 7.03 & 218& Developed approach solved via \cite{104} & 3.41 \\
                       \hline
                      LASSO& 0.35 & 1.71 & 4.76 & 44.32& Developed approach solved via \cite{104}& 1.53\\
                                          
                     \hline
                     \end{tabular}
                     \end{center}
                      \end{table*}
\begin{figure}[t!]
          \includegraphics[scale=0.54]{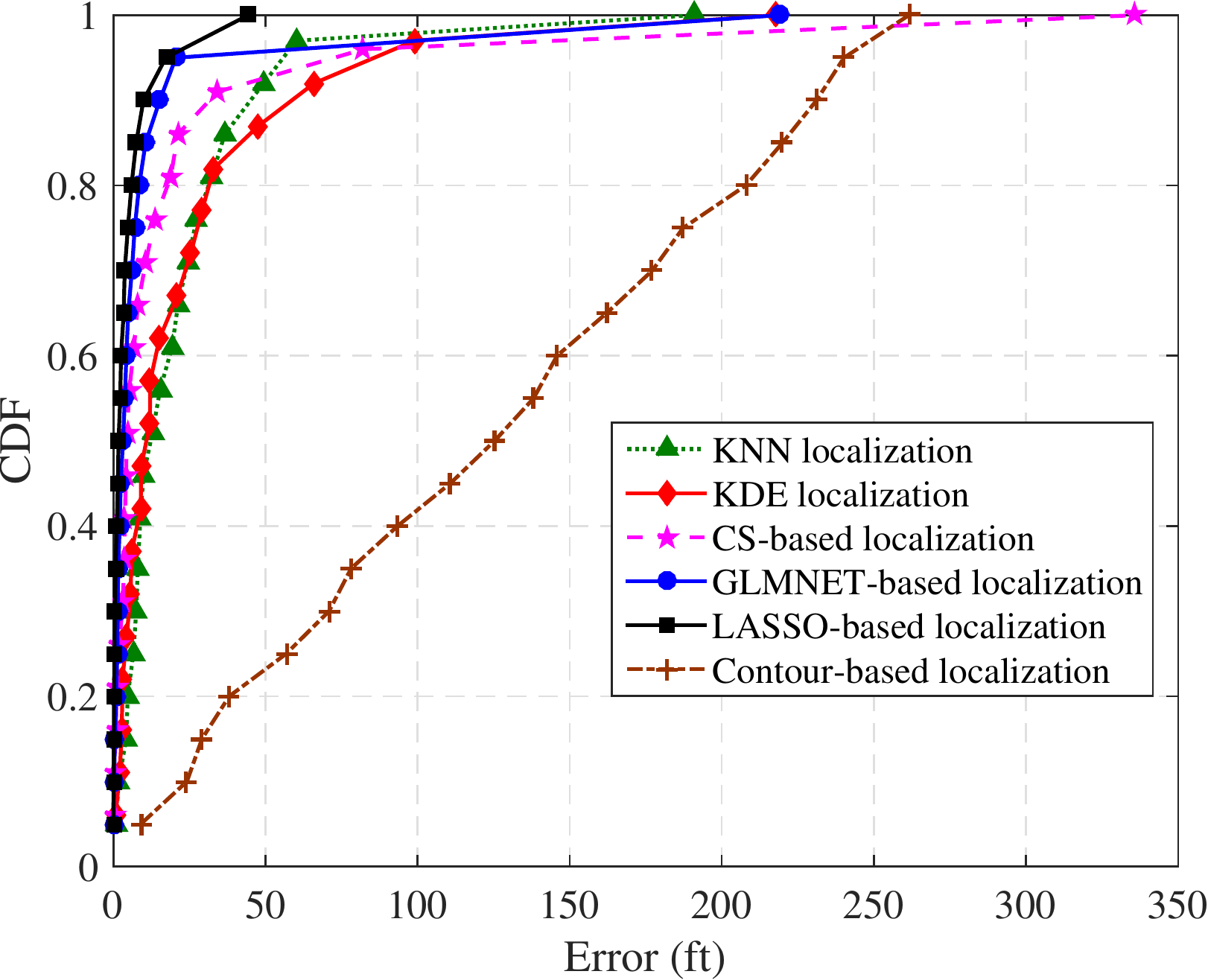}
               \centering

               \caption{The cumulative position error distribution for KNN, KDE, CS-based, Contour-based, GLMNET and LASSO-based method.}
               \label{figure2}
            \end{figure}
\par The average localization error is merely a first-order statistics for assessing a localization scheme. In fact, one should consider the spread and frequency of errors as well. To this end, we evaluate the empirical Cumulative Distribution Function (CDF) of the position estimate errors. Fig. \ref{figure2} depicts the CDF of the localization error for the two proposed schemes along with some well-known localization schemes when 10 APs are used. The LASSO shows an error of not more than 44.32 ft while GLMNET gives 92\% of the errors in this range. Approximately,  2\% of the errors for CS-localization is more than 160 ft. The CS-based localization gives 98\% of pose estimate errors less than 102 ft. The Contour-based localization method does not perform well in our environment. A possible reason is that the path loss model in office environments is complicated due to several layers of the walls, however, the method performs better in open areas such as malls and airports \cite{99}.
\begin{figure}[t!]
          \includegraphics[scale=0.542]{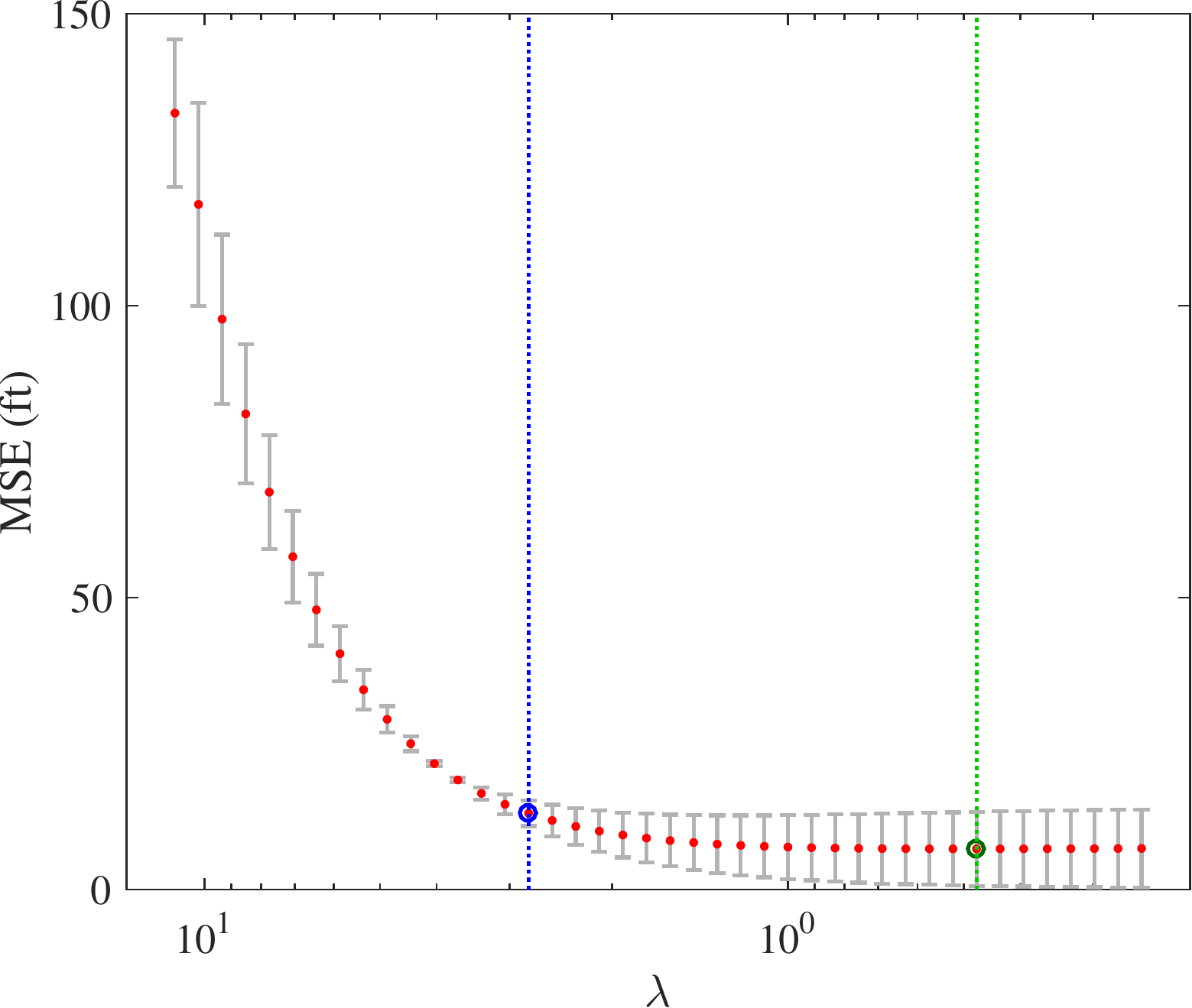}
               \centering
               \caption{The two-fold cross validation for finding  $\lambda$ with the minimum MSE for LASSO.}
               \label{figure6}
            \end{figure}
            
            \begin{figure}[t!]
                   \includegraphics[scale=0.48]{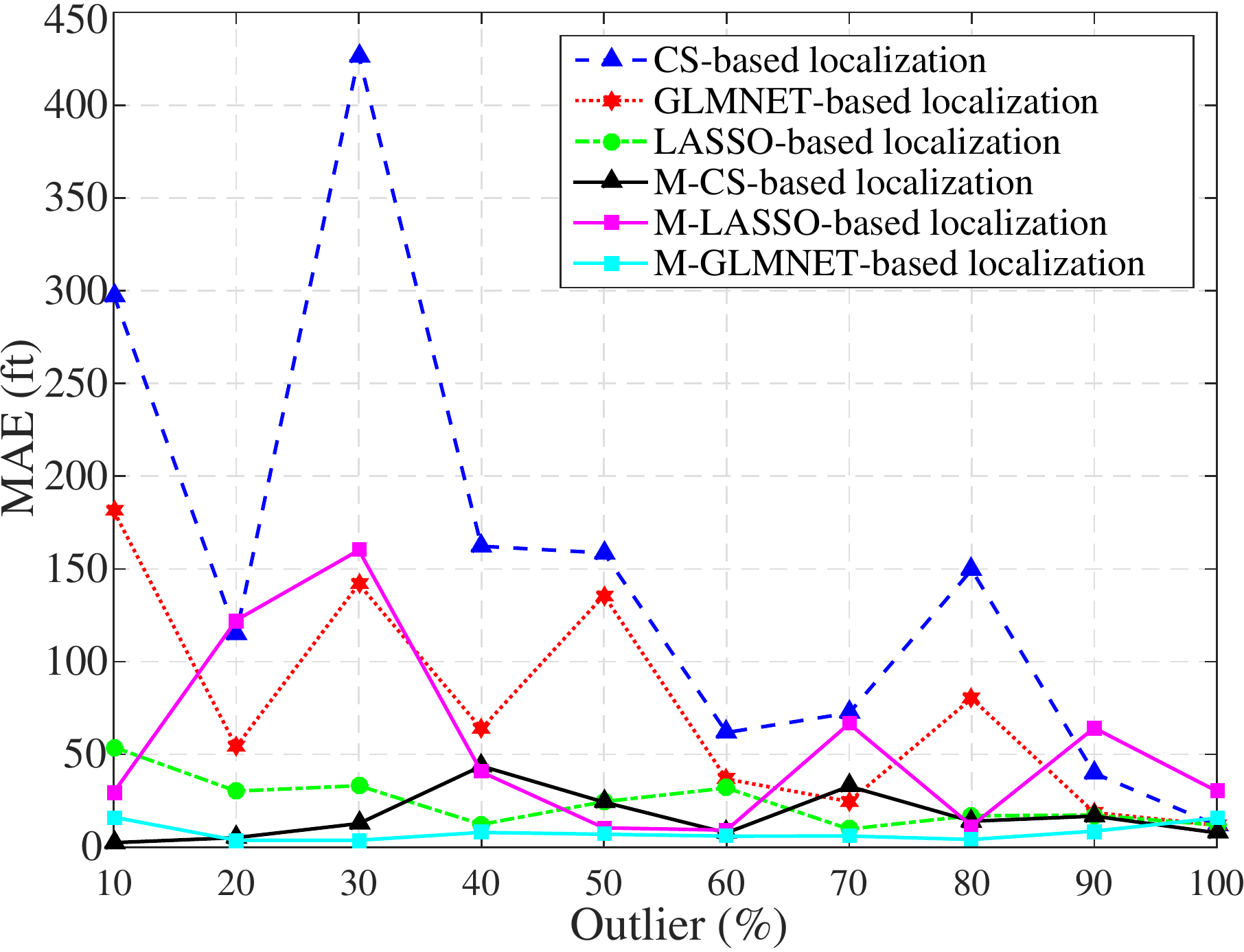}
                        \centering
                        \caption{The outlier detection capabilities of the proposed methods with an increasing rate of APs containing outliers.}
                        \label{figure7}
                     \end{figure}

\par The percentiles of the position estimate errors are shown in Table 2 for the proposed methods and compared with replicated conventional algorithms. The CS has the least localization errors among the compared conventional techniques. Considering the 50\% percentile, the GLMNET and LASSO exhibit respectively 30\% and 61\% accuracy improvement relative to the CS method, and thus, outperform all others. The table also shows the running time of the online localization phase which conveys the computational efficiency of the methods. WKNN delivers the least running time. The GLMNET and LASSO, with $3.41$ ms and  $1.53$ ms running times respectively, provide smaller localization errors compared to other methods but only GLMNET requires slightly longer running time. Comparing with a popular non-fingerprinting technique, EZ (which is a model-based approach with $11.48$ ft error), one can note that it is designed to simplify the deployment and is less accurate than fingerprinting approaches, including the proposed methods. Comparison with the methods that use additional infrastructure components such as Centaur with acoustic ranging tools is out of scope of this paper.  

\par LASSO and GLMNET localization schemes need tuning of the parameters $\lambda$, $\alpha$, and $\mu$. The usual way is to use cross-validation (CV) \cite{52} that computes the prediction error to guide the choice of tuning parameters. CV searches over a range of values so that the Mean Square Error (MSE) of residuals becomes optimal. Fig.\ref{figure6} shows a one-dimensional cross-validation over $\lambda$ for LASSO. Also, a set of training data can be utilized to tune these parameters.
\par Fig.\ref{figure7} shows the average localization error for the CS-based, LASSO-based, and GLMNET-based localization in two different scenarios when the APs contain outliers. In the first scenario, the user's position is estimated through solving the localization problem (\ref{eq4}), (\ref{eq17}), and (\ref{eq18}). The second scenario estimates the user's position through jointly detecting the outlier and estimating the user's position (\ref{eq26}), (\ref{eq27}), and (\ref{eq28}). In both scenarios 10 APs have been used and the number of APs containing the outlier is increasing. The results show that in the first scenario, the CS-based localization introduces significant errors while the LASSO is more robust to outliers. However, the M-CS method has the least errors if the number of APs containing outliers is low. But, as the number of outlier-contaminated APs increases, M-GLMNET is more robust and keeps the positioning error smaller than the other approaches.

\section{Conclusions}   

In this paper, we have proposed two novel sparse recovery algorithms for indoor WLAN fingerprinting positioning. We showed that a simple clustering algorithm can efficiently cluster the RPs when the clustering scheme includes overlapped clusters which are connected through neighbor clusters. The two proposed localization approaches, LASSO and GLMNET, are able to provide a significant lower localization error as they also minimize the residuals between the radio map and online measurement. In addition, rather than being negatively affected by the correlated radio map fingerprints, they take benefit of the similarity between RSS readings and do not need orthogonalization pre-processing. We also showed that if the APs contain outliers, which is probable, the positioning error would be high. Hence, through augmenting the proposed method, our methods can jointly detect, modify the outlier, and estimate the user's position. The experimental results on a real set of data showed that the proposed methods can efficiently and effectively localize the mobile user's through WLAN signals.

\section{Future Work}
Indoor localization research is very extensive and this paper addressed an advanced solution which exploits only fingerprinting measurements. Comparison of many diverse localization techniques is hindered by the lack of standardized representative data that can be used for fair comparisons. The authors plan to create an open repository of their data that can be used by the community for comparative studies.
\par The radio map addressed in this work has four orientations. The proposed localization scheme integrates the available information from all orientations as described. However, there might be alternative formulations for handling different orientations, which is of great interest for further studies. 
\par In addition, the experimental results in the previous section showed that the fine localization algorithms represent fluctuations in localization accuracy for small number of APs. An interest of our future research is to find the causes of this phenomena and provide more stable localization mechanism with smaller number of APs.
\section{Acknowledgment}
The authors would like to thank the anonymous reviewers for their constructive comments which have improved the quality of this paper. 
\ifCLASSOPTIONcaptionsoff
  \newpage
\fi

\bibliographystyle{IEEEtran}
\bibliography{mybib}
\newpage
\begin{IEEEbiography}[
{\includegraphics[width=1in,height=1.25in,clip,keepaspectratio, angle =0 ]{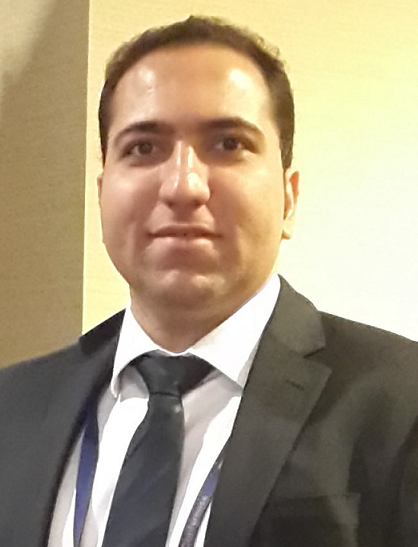}}
]{Ali Khalajmehrabadi} received his B.Sc. from Babol Noshirvani University of Technology, Iran, in 2010 and M.Sc. from University Technology Malaysia (UTM), Malaysia, in 2012 with the best graduate student award. He is currently a Ph.D. candidate in the Department of Electrical and Computer Engineering, the University of Texas at San Antonio (UTSA). His research interests include indoor localization and navigation systems, collaborative localization,  and  Global Navigation Satellite System (GNSS). He is a student member of IEEE and Institute of Navigation (ION). 
\end{IEEEbiography}
\vspace{-3 mm}
\begin{IEEEbiography}[
{\includegraphics[width=1in,height=1.25in,clip,keepaspectratio, angle =0 ]{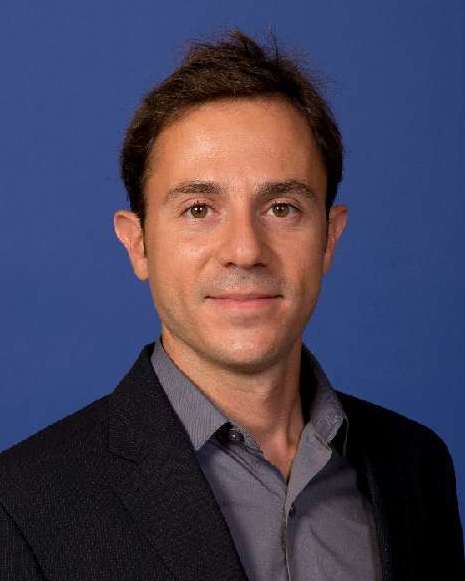}}
]{Nikolaos Gatsis}
 received the Diploma degree in Electrical and Computer Engineering from the University of Patras, Greece, in 2005 with honors. He received the M.Sc. degree in Electrical Engineering in 2010, and the Ph.D. degree in Electrical Engineering with minor in Mathematics in 2012, both from the University of Minnesota. He is currently an Assistant Professor with the Department of Electrical and Computer Engineering at the University of Texas at San Antonio. His research interests lie in the areas of smart power grids, communication networks, and cyberphysical systems, with an emphasis on optimal resource management. Prof. Gatsis co-organized symposia in the area of Smart Grids in IEEE GlobalSIP 2015 and IEEE GlobalSIP 2016. He also served as a Technical Program Committee member for symposia in IEEE SmartGridComm from to 2013 through 2016, and in GLOBECOM 2015.
\end{IEEEbiography}
\vspace{-3 mm}
\begin{IEEEbiography}[
{\includegraphics[width=1in,height=1.25in,clip,keepaspectratio, angle =0 ]{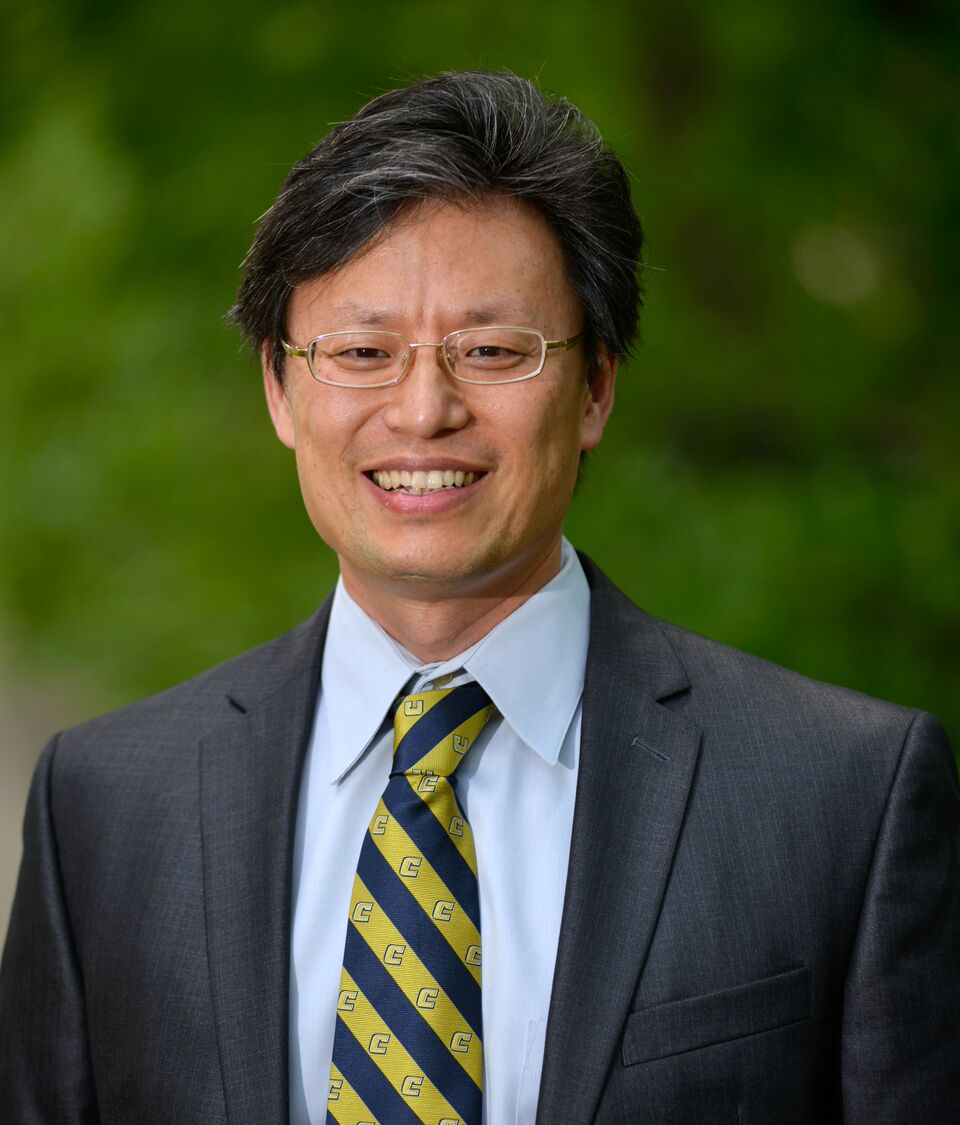}}
]{Daniel J. Pack}
received the Bachelor of Science degree in Electrical Engineering, the Master of Science degree in Engineering Sciences, and the Ph.D. degree in Electrical Engineering from Arizona State University, Harvard University, and Purdue University, respectively.  He is the Dean of the College of Engineering and Computer Science at the University of Tennessee Chattanooga (UTC).  Prior to joining UTC, he was Professor and Mary Lou Clarke Endowed Chair of the Electrical and Computer Engineering Department at the University of Texas, San Antonio, after serving as Professor of Electrical and Computer Engineering at the United States Air Force Academy, CO.  His research interests include unmanned aerial vehicles and intelligent control. He is a member of Eta Kappa Nu, Tau Beta Pi, IEEE, AIAA and ASEE and serves an Associate Editor for IEEE Systems Journal.  

\end{IEEEbiography}

\begin{IEEEbiography} [

{\includegraphics[width=1in,height=1.25in,clip,keepaspectratio, angle =0 ]{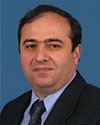}}
]{David Akopian} (M’02-SM’04) is a Professor at the University of Texas at San Antonio (UTSA). Prior to joining UTSA he was a Senior Research Engineer and Specialist with Nokia Corporation from 1999 to 2003. From 1993 to 1999 he was a researcher and instructor at the Tampere University of Technology, Finland, where he received his Ph.D. degree in electrical engineering in 1997. Dr. Akopian’s current research interests include digital signal processing algorithms for communication and navigation receivers, positioning, dedicated hardware architectures and platforms for software defined radio and communication technologies for healthcare applications. He authored and co-authored more than 30 patents and 140 publications. He served in organizing and program committees of many IEEE conferences and co-chairs annual SPIE Multimedia on Mobile Devices conference. His research has been supported by National Science Foundation, National Institutes of Health, USAF, US Navy and Texas foundations.
\end{IEEEbiography}




\end{document}